\documentclass{article}
\usepackage{fullpage}

\title{A Super-Grover Separation Between Randomized and Quantum Query Complexities}
\author{Shalev Ben-David\thanks{Computer Science and Artificial Intelligence Lab,
				Massachusetts Institute of Technology. shalev@mit.edu}}

\begin{document}
\maketitle

\begin{abstract}
We construct a total Boolean function $f$ satisfying
$R(f)=\tilde{\Omega}(Q(f)^{5/2})$, refuting the long-standing
conjecture that $R(f)=O(Q(f)^2)$ for all total Boolean functions.
Assuming a conjecture of Aaronson and Ambainis about optimal quantum speedups for partial functions,
we improve this to $R(f)=\tilde{\Omega}(Q(f)^3)$.
Our construction is motivated by the G\"o\"os-Pitassi-Watson function
but does not use it.
\end{abstract}

\section{Introduction}

The deterministic query complexity of a function $f:\{0,1\}^N\to\{0,1\}$
is the minimum, over all deterministic algorithms,
of the worst-case number of queries to the bits of
an input $x\in\{0,1\}^N$ the algorithm requires to determine $f(x)$.
This value is denoted by $D(f)$.
The randomized query complexity, denoted by $R(f)$, is defined similarly,
except the algorithm is allowed to use randomness and is only required
to correctly determine $f(x)$ with probability at least $2/3$ for all $x$.
The quantum query complexity $Q(f)$ is analogous to $R(f)$, except the
algorithm is allowed quantum queries.

It is clear that $Q(f)\leq R(f)\leq D(f)$.
Beals, Buhrman, Cleve, Mosca, and de Wolf 
\cite{Beals01} showed that for all total functions $f$, $D(f)=O(Q(f)^6)$.
In particular, this implies that $R(f)=O(Q(f)^6)$.
For a summary of query complexity relations, see \cite{Buhrman02}.

Until recently the only known gap between $D(f)$ and $Q(f)$ was quadratic.
It was conjectured that this is tight.
In a recent breakthrough, Ambainis, Balodis, Belovs, Lee, Santha, and Smotrovs
\cite{Pointer} provided a total function exhibiting a degree $4$ separation
between $D(f)$ and $Q(f)$, refuting this conjecture.
This function was a variant of the G\"o\"os-Pitassi-Watson (GPW) function,
introduced in \cite{GPW}.
However, this function had only a quadratic gap between
$R(f)$ and $Q(f)$, as its quantum speedup was based only on Grover search,
which fundamentally provide at most a quadratic speedup.
This has led to the suggestion that the $R(f)=O(Q(f)^2)$ conjecture
might still hold for total Boolean functions.

In this work, we provide a total function exhibiting super-Grover
(degree $2.5$) quantum speedup over randomized algorithms. By assuming a conjecture
of Aaronson and Ambainis \cite{Forrelation}, we improve this separation to degree $3$.
Our construction is not directly based on the GPW function, and is substantially
different from those the GPW variants in \cite{Pointer}. However, it is similar to GPW
in several ways, including the use of pointers and depth-$2$ and-or trees.

\section{Preliminaries}

The following two functions will be used in our construction.

\textbf{Forrelation}: The \textsc{Forrelation}
function was introduced by Aaronson and Ambainis \cite{Forrelation}.
It provides the largest known gap between randomized and quantum
query complexities in the promise setting, with $Q(f)=O(1)$ and
$R(f)=\tilde{\Omega}(\sqrt{m})$ (where $m$ is the size of the input).
A variant of \textsc{Forrelation}
is conjectured to satisfy $Q(f)=O(\log m)$ and $R(f)=\tilde{\Omega}(m)$.

\textbf{Or-And}: This function is an \textsc{Or} on $m$ bits
composed with $m$ copies of \textsc{And} on $m$ bits. In other words,
its input is an $m$-by-$m$ matrix, and the value of the function is $1$
if and only if there is an all-$1$ column in the matrix. The randomized query
complexity of \textsc{Or-And} is $\Theta(m^2)$, and the quantum
query complexity is $\Theta(m)$. In addition, note that an \textsc{Or-And}
instance always has a certificate of size $m$: either an all-$1$ column,
or else a $0$ in each column. The final property we will use is that
if a quantum algorithm is given pointers to the $m$ bits in a
certificate for \textsc{Or-And}, it can verify the certificate
using $O(\sqrt{m})$ queries.

\section{Constructing the Function}

\subsection{Sketch of Construction}
The idea of the construction is as follows.
The \textsc{Forrelation} function
provides a large quantum speedup in the promise setting, requiring
$\tilde{\Omega}(\sqrt{m})$ randomized queries but only $O(1)$
quantum queries. We wish to turn it into a total function.
To do this, we compose \textsc{Forrelation} with \textsc{Or-And}; that is,
we replace each bit of the \textsc{Forrelation} function
with an $m$ by $m$ copy of \textsc{Or-And}.
This slows down both the randomized and quantum algorithms.
Any randomized algorithm now uses $\tilde{\Omega}(m^{5/2})$ queries,
while a quantum algorithm can use $\tilde{O}(m)$.
For lack of a better name, we will refer to this composed function as \textsc{Forandlation}.

We then use the solution of the \textsc{Forandlation} instance
to find a ``cheat sheet'' that links to the
certificates of all the \textsc{Or-And} instances; using this
cheat sheet, a quantum algorithm can verify that the input
satisfied the \textsc{Forrelation} promise without solving the \textsc{Or-And}
instances. In fact, it can do the verification using $\tilde{O}(m)$ queries.
We can then make the value of the function equal to $0$ if the verification
fails, which turns the function total.

How do we use the \textsc{Forandlation} instance to find such a cheat sheet,
while ensuring that a randomized algorithm cannot find it?
We do this by assuming that the input contains not just the \textsc{Forandlation}
instance, but also an array of size (say) $m^{10}$ full of candidate sheets.
In addition, instead of one \textsc{Forandlation} instance we will need $10\log m$
\textsc{Forandlation} instances. We will use the answers of these \textsc{Forandlation} instances as the index
of a sheet in the array, and assert that this is the cheat sheet.
A randomized algorithm cannot find this cheat sheet,
but a quantum algorithm can.

Once again, if anything goes wrong - if the cheat sheet does not link to proper
certificates, or if the certified \textsc{Or-And} answers don't make up instances
in the promise of \textsc{Forrelation} - we set the value of the function to $0$.
In all other cases, the value of the function will be a special bit on the cheat sheet.

\subsection{Formal Construction}

Let $x$ be the input. We define the value of our function $g$
on $x$ as follows.

Let $m$ be (say) the $20$-th root of the input size.
We will interpret the first $10m^3 \log m$ bits of $x$ as
$10\log m$ instances of \textsc{Forandlation} (each containing $m^3$ bits).
The next $O(m^{12}\log^2 m)$ bits of $x$ will be interpreted as an array of size $m^{10}$
whose entries have $O(m^2 \log^2 m)$ bits each. The rest of the input will be ignored.

If the \textsc{Forandlation} instances don't all satisfy their promise,
we set $g(x):=0$. Otherwise,
the solution to the \textsc{Forandlation} instances will be used as an index of
the array. The entry of the array corresponding to that index will be called the cheat sheet.

The first $10m\log m$ bits of the cheat sheet will be interpreted as
the solutions to all the \textsc{Or-And} instances inside the
$10\log m$ \textsc{Forandlation}s, in order. If these bits are not equal to
the \textsc{Or-And} values, we set $g(x):=0$.

The next $O(m^2\log^2 m)$ bits of the cheat sheet will be interpreted
as $10m^2\log m$ pointers to bits in the \textsc{Forandlation} part of the input.
These pointers should correspond to certificates for each of the $10m\log m$
\textsc{Or-And} instances (all the certificates have size $m$).
If they do not, we set $g(x):=0$.

Finally, if the value of $x$ has not yet been set to $0$, we set it
to the bit of the cheat sheet which immediately follows the aforementioned
pointers.

\section{Quantum Upper Bound}

We provide an algorithm for evaluating $g$ using $\tilde{O}(m)$
queries. For simplicity, we will ignore log factors in our analysis, which allows us
to ignore amplification concerns.

First, note that \textsc{Forandlation} can be solved by a quantum
algorithm using $O(m)$ queries. This is because \textsc{Forandlation}
is a composition of \textsc{Forrelation}, which requires $O(1)$
queries, and \textsc{Or-And}, which requires $O(m)$ queries.

Our quantum algorithm will begin by evaluating the $10\log m$
\textsc{Forandlation} instances in the input, using $\tilde{O}(m)$ queries.
It will then go to the designated entry of the array, and read the first
$10m\log m$ bits (the ones representing the answers to the
\textsc{Or-And} instances). This takes an additional $\tilde{O}(m)$
queries. The algorithm will then verify that the $10m\log m$ 
bits it read satisfy the promises of $10\log m$ \textsc{Forrelation} instances.
If a promise isn't satisfied, the algorithm will return $0$.

Next, the quantum algorithm will use the $10m^2\log m$
pointers to Grover-search for an invalid certificate.
There are $10m\log m$ certificates to check, and
each one takes $O(\sqrt{m})$ quantum queries to verify,
so this Grover search takes $\tilde{O}(m)$ queries.
If a bad certificate is found, the algorithm returns $0$.

Finally, the algorithm will query the remaining bit of the
array entry and return its value.

This algorithm uses a total of $\tilde{O}(m)$ queries.
Its correctness is easy to see: the algorithm only returns
$1$ if all the \textsc{Forandlation} promises hold and
the cheat sheet successfully certifies this fact.

\section{Randomized Lower Bound}

\subsection{Forandlation Lower Bound}

We show that a randomized algorithm will take at least $\tilde{\Omega}(m^{5/2})$
queries to solve a \textsc{Forandlation} instance consisting of $m^3$ bits.

Let $A$ be any randomized algorithm for this problem.
We turn $A$ into an algorithm $B$ for solving \textsc{Forrelation} on $m$ bits.
Consider any input $x$ for \textsc{Forrelation} on $m$ bits.
Choose $m$ random $m$-by-$m$ matrices with the property that each
column of the matrix has exactly one $0$. Then, for $i=1,2,\dots,m$,
pick a random $0$ in the $i$-th matrix and replace it with $x_i$.
This construction turns the input $x$ for \textsc{Forrelation}
into an input $x^\prime$ for \textsc{Forandlation} which has the same value.

Now, we get the algorithm $B$ to simulate $A$ on $x^\prime$. Whenever $A$ queries a bit of $x$,
$B$ queries the corresponding input of $x$ and gives $A$ the answer
 Once $A$ terminates,  $B$ returns the answer it gives.

We'll show that a randomized algorithm must use $\tilde{\Omega}(m^2)$ queries to find the special $x_i$
bit in a matrix. This will imply that if $A$ succeeds with probability at least $2/3$ after at most $k$ queries,
then $B$ succeeds after $\tilde{O}(k/m^2)$ queries to $x$. Since \textsc{Forrelation}
requires $\tilde{\Omega}(\sqrt{m})$ queries, it follows that $A$ requires
$\tilde{\Omega}(m^{5/2})$ queries. If \textsc{Forrelation} were replaced by a function
with randomized query complexity at least $\tilde{\Omega}(m)$, this would become
$\tilde{\Omega}(m^3)$ instead.

It remains to show that a randomized algorithm must use $\tilde{\Omega}(m^2)$
queries to find $x_i$. We can think of $x_i$ as a random marked zero in a matrix that has
exactly one zero randomly placed in each column (the rest of the matrix is filled with ones).
To find the marked zero with probability at least $1/2$, a randomized algorithm
must find at least half the zeros. Now, if a randomized algorithm made only
$O(m/\log m)$ queries in a given column, then its chance of finding a zero in that column
is at most $O(1/m)$. By the union bound, the randomized algorithm will only discover
a zero in columns in which it queried at least $\Omega(m/\log m)$ bits.
Since we require the randomized algorithm to find $m/2$ zeros, the total number
of queries it must make is at least $\tilde{\Omega}(m^2)$. This
completes the argument.

\subsection{Lower Bound for $g$}

We now show a lower bound for the randomized query complexity of our function $g$.
Let $A$ be a randomized algorithm for $g$ that uses
at most $k$ queries. Consider giving $A$ an input which satisfies the promises
of all the \textsc{Forandlation} instances, but which has an empty array (that is,
the array contains only zeros). Then the value of $g$ on this instance is $0$, so
$A$ outputs $0$ after $k$ queries with probability at least $2/3$.
We can amplify this to $99/100$ and increase $k$ by only a constant factor.

Now, if we instead give $A$ an instance which is identical except
that the appropriate cell of the array contains a valid cheat sheet,
then $A$ must output $1$ with probability at least $99/100$.
This means a bit of the appropriate cell was queried with high probability.

In other words, when we run $A$ with an empty array,
we get a list of at most $k$ spots in the array where the cheat sheet should be
(this list is simply the cells of the array that are queried by this randomized algorithm).
This list contains the correct location for the cheat sheet with high probability.

We can think about the indices of the array as binary strings.
The algorithm $A$ gives us a set $S$ of at most $k$ binary strings
such that the string of answers to the \textsc{Forandlation}
instances is in $S$ (with high probability).

Let $\mathcal{D}$ be a hard distribution over the inputs of \textsc{Forandlation},
so that any randomized algorithm must make $\tilde{\Omega}(m^{5/2})$
queries to evaluate \textsc{Forandlation} on a random input sampled from $\mathcal{D}$.
We can split $\mathcal{D}$ into a distribution $\mathcal{D}^0$ over $0$-inputs
to \textsc{Forandlation} and a distribution $\mathcal{D}^1$ over $1$-inputs
such that distinguishing between these distributions takes $\tilde{\Omega}(m^{5/2})$
queries for a randomized algorithm.

Now, consider giving $A$ $10\log m$ inputs to \textsc{Forandlation}
that are each sampled from $\mathcal{D}^0$. The algorithm $A$
produces a set $S$ of binary strings, and with high probability,
the all-$0$ string is in $S$.

Assuming $k < m^5$, there is some binary string of length $10\log m$
whose probability of being in $S$ is less than $1/m^5$. Let $t$
be such a string. Consider a sequence of strings starting with the all-$0$
string and ending with $t$, such that any two consecutive strings
$x$ and $y$ in the sequence differ in exactly one bit, with $y$ having
one more $1$ than $x$. This sequence will have length $O(\log m)$.

For each string $x$ in the sequence, we form an input to $A$ by concatenating
samples from $\mathcal{D}^{x_i}$ for $i=1,2,\dots,10\log m$, and adding a blank array.
This gives us $O(\log m)$ distributions over inputs to $A$.
By construction, $A$ distinguishes the first distribution (based on the all-zero string)
from the last distribution (based on $t$), since the former will have very small probability
of including $t$ in $S$ while the latter will have high probability.

From this it follows that there is some consecutive pair of strings in this sequence
that $A$ distinguishes with probability at least $\Omega(1/\log m)$.
However, distinguishing a consecutive pair can be used to distinguish
$\mathcal{D}^0$ and $\mathcal{D}^1$, which is as hard as \textsc{Forandlation}.
It follows that $A$ uses at least $\tilde{\Omega}(m^{5/2})$ queries, as desired.
The same proof shows a lower bound of $\tilde{\Omega}(m^3)$
if there is a function $f$ with $Q(f)=\tilde{O}(\log m)$ and $R(f)=\tilde{\Omega}(m)$.

\section{Final Remarks}

In this proof, we used several specific properties of the \textsc{Or-And} function,
but no properties of \textsc{Forrelation} other than its randomized and quantum query complexities.
This means the same proof technique will extend to functions other than \textsc{Forrelation},
and may be used to provide separations between other query complexity measures.
It might also be interesting to try to combine this technique with the
G\"o\"os-Pitassi-Watson variants of \cite{Pointer}.

\section*{Acknowledgements}
I would like to thank Scott Aaronson and Robin Kothari for checking an early draft
of this result.

\bibliographystyle{plain}


\end{document}